# A comprehensive first-principles analysis of phonon thermal conductivity and electron-phonon coupling in different metals

Zhen Tong[1,2], Shouhang Li[1], Xiulin Ruan[2] and Hua Bao*[,1]

[1]University of Michigan-Shanghai Jiao Tong University Joint Institute, Shanghai Jiao Tong University, Shanghai 200240, China.

[2]School of Mechanical Engineering and the Birck Nanotechnology Center, Purdue University, West Lafayette, Indiana 47907-2088, USA

**Abstract** Separating electron and phonon thermal conductivity components is imperative for understanding the principle thermal transport mechanisms in metals and highly desirable in many applications. In this work, we predict the mode-dependent electron and phonon thermal conductivities of 18 different metals at room-temperature from first-principles. Our first-principles predictions in general agree well with experimental data. We find that the phonon thermal conductivity is in the range of 2 - 18 W/mK, which accounts for 1% - 40% of the total thermal conductivity. It is also found that the phonon thermal conductivities in transition metals and transition-intermetallic-compounds (TICs) are non-negligible compared to noble metals due to their high phonon group velocities. Besides, the electron-phonon coupling effect on phonon thermal conductivity in transition metals and intermetallic compounds is stronger than that of nobles, which is attributed to larger electron-phonon coupling constant with a high electron density of state within Fermi window and high phonon frequency. The noble metals have higher electron thermal conductivities compared to transition metals and TICs, which is mainly due to the weak electron-phonon coupling in noble metals. It is also shown that the Lorenz ratios of transition metals and transition-intermetallic-compounds hold larger deviations from the Sommerfeld value $L_0$=2.44×10$^{-8}$ WΩK$^{-2}$. We also find the mean free paths (MFPs) for phonon (within 10 nm) are smaller than those of electron (5 - 25 nm). The electrical conductivity and electron thermal conductivity are strongly related to the MFPs of electron.

* Author to whom correspondence should be addressed. *E-mail address*: hua.bao@sjtu.edu.cn (H.Bao).

## 1. Introduction

There have been numerous experimental measurements and theoretical analysis in order to understand the thermal transport in metals ever since the early of the twentieth century[1–10]. Nowadays, it has been generally believed that free electrons have dominant contribution to the thermal transport in metals, while the phonons have less contribution. In most applications, only the total thermal conductivity of metals is needed, so it is unnecessary to separate electron and phonon thermal conductivity. Recently, there has been growing interest in quantifying phonon heat conduction in metals, primarily driven by recent research advances in a variety of electron-phonon nonequilibrium energy transfer processes, for example, thermal transport across metal-dielectric interface[11], laser manufacturing and laser heating[12], heat-assisted magnetic recording devices[13], *etc*. In addition, resolving the size effect of metal nanostructure also requires the quantification of electron and phonon thermal conductivity and mean free path of metals[14–17]. Therefore, how to separate the electron and phonon thermal conductivity components of metals becomes an important problem, which prompts researchers to carry out various experimental and theoretical works.

In most pure metals, the electron thermal conductivity ($\kappa_e$) is believed to be much larger than the phonon thermal conductivity ($\kappa_p$). Also, it has been well known that the electron thermal conductivity of metals is proportional to the electrical conductivity, or the Wiedemann-Franz law[18]. Therefore, to obtain a simple estimation of electron thermal conductivity, one can perform electrical conductivity measurement, for example, 4-probe resistivity measurement[19] and then employ Wiedemann-Franz law[18] to obtain the electron thermal conductivity. To further obtain the phonon thermal conductivity, the total thermal conductivity can be measured and the electron thermal conductivity $\kappa_e$ is estimated from electrical conductivity, and then the difference can be attributed to phonon contribution[1,20,21]. To apply the Wiedemann-Franz law, it generally needs a correct Lorenz number, which is usually taken as the Sommerfeld value[1,10]. However, it has been well known that the Sommerfeld value only holds when elastic scattering prevails, which is generally limited to low or high temperature[4,22]. Because phonon thermal conductivity is a relatively small fraction, even a small deviation of Lorenz number can lead to large uncertainty in the phonon thermal conductivity evaluation. In order to more accurately obtain phonon thermal conductivity, a few other experimental methods have been implemented, including alloying method[5,6], superconducting method[5,23], and magnetothermal method[6,24]. However, these methods are either very complicated to conduct or limited to extremely low temperature. Therefore, the accuracy of experimentally measured $\kappa_p$ is still uncertain and even the room-temperature values are only available for few metals[5–9].

From the theoretical side, in order to investigate the phonon thermal conductivity of metals, the general strategy is to first estimate phonon thermal conductivity considering the phonon-phonon scattering. Some of the early efforts include, Leibfried and Schlömann[25] model, Klemens[4] model, and Slack[26] equation. All these analytical models only involve the phonon-phonon scattering, which neglect the phonon-electron scattering in metals[27,28]. By adding the phonon-electron scattering rate, Klemens[10] proposed the formula of phonon thermal conductivity by assuming that the long wavelength phonon modes interact with free electrons and then concluded that the phonon thermal conductivity of pure metals are in range of 3 to 10 W/mK. Stojanovic *et al.*[29] developed an analytical expression of phonon thermal conductivity for metal nanostructures with the assumption of isotropic properties of the material, nearly free electron for electrons and Debye approximation[30] for phonons. On the other hand, the expression of electron thermal conductivity at low and high temperature for monovalent metals was firstly derived by Wilson[31] with the assumption that only the longitudinal phonon modes interact with electrons. Makinson[1] promoted the expressions of electron thermal conductivity for both high and low temperatures by assuming that the phonons with different polarization interact with electrons to the same extent. These theoretical treatments significantly advance the understanding of thermal transport in metals. However, large uncertainty will be induced if using these models to obtain quantitative phonon and electron thermal conductivity values.

On the other hand, recent advances in numerical methods have enabled more accurate prediction of phonon thermal conductivity in metals. For example, the molecular dynamic method was used to predict the phonon thermal conductivity of metals[32]. However, molecular dynamics has significant limitation, not only because the accuracy of force fields is questionable, but also because empirical potential is unavailable for most materials. In contrast, by first-principles method, it is possible to extract the electron-phonon coupling matrix element, and then the mode-resolved electron and phonon transport properties can be obtained by combining with Boltzmann transport equation[33–36]. Therefore, this method can be quite useful to obtain a reliable phonon thermal conductivity of metals. First-principles method can be ideally applied to any material. The major limitation is that very dense **k**-mesh and **q**-mesh used for Brillouin zone integration are needed to obtain accurate results, which requires extremely high computational cost. In recent years, there are only a handful of first-principles calculations[27,37–39] carried out to predict the thermal conductivity of metals. For example, Jain *et al.*[37] predicted the electron and phonon thermal conductivity of Au, Ag and Al including the electron-phonon scattering rate by using dense **k**-mesh (80×80×80) and **q**-mesh (32×32×32) interpolation. Wang *et al.*[27] calculated the phonon thermal conductivity of metals including aluminum (Al), noble metals (Au, Ag, Cu) and transition metals (Pt, Ni), but the accuracy of their calculated values could be limited due to the relatively coarse mesh of **k**-points (16×16×16) and **q**-points (16×16×16) used in the Brillouin zone integration. In our previous work[38], the intermetallic compound NiAl and $Ni_3Al$ were considered. These advances are important in that one can finally obtain relatively reliable values of phonon thermal conductivity in metals. However, these first-principles simulations are scattered to only a few types of metals, and the data may not be directly comparable due to the difference in their simulations, for example, the pseudopotential and Brillouin zone integration technique. Therefore, a comprehensive analysis of the phonon thermal conductivity in different types of metals is highly desirable in order to obtain more general conclusions of heat conduction in metals.

In this work, a series of first-principles calculations are carried out to predict the mode-dependent electron and phonon thermal conductivity of 18 different metals, which include noble metals, alkali-earth metals, transition metals, transition-intermetallic-compounds (TICs) and noble-intermetallic-compounds (NICs). The phonon thermal conductivities are calculated by considering both phonon-phonon (*p*-*p*) and phonon-electron (*p*-*e*) interactions, and the effect of phonon-electron scattering on the phonon thermal conductivity is carefully discussed. In addition, the electron thermal conductivity is evaluated by considering electron-phonon (*e*-*p*) scattering and the Lorenz numbers, as well as the mean free paths for both phonon and electron of all 18 metals are calculated.

## 2. Methods and simulation details

### 2.1. Methods

#### 2.1.1. Phonon thermal conductivity

Combining the Boltzmann transport equation (BTE) and the Fourier's law[40], the phonon thermal conductivity tensor can be calculated as

$$\kappa_p^{\alpha\beta} = \sum_{\lambda} c_{v,\lambda} v_{\lambda}^{\alpha} v_{\lambda}^{\beta} \tau_{\lambda}^{p}, \qquad (1)$$

where $\lambda=(\mathbf{q},\nu)$ denotes the phonon mode with polarization $\nu$ and wave vector $\mathbf{q}$, $c_{v,\lambda}$ is the volumetric specific heat, $v_{\lambda}^{\alpha}$ and $v_{\lambda}^{\beta}$ is the $\alpha$ and $\beta$-component of the phonon group velocity vector $v_{\lambda}$, and $\tau_{\lambda}^{p}$ is the phonon relaxation time. The phonon volumetric specific heat can be obtained by using the Bose-Einstein statistics as $c_{v,\lambda} = \frac{\hbar\omega_\lambda}{V}\frac{\partial n_\lambda}{\partial T}$, where $n_\lambda$ is the Bose-Einstein distribution function and $V$ is the volume of the primitive cell. The group velocity can be obtained by $v_{\lambda,\alpha} = \frac{\partial \omega_\lambda}{\partial \mathbf{q}}$. The phonon relaxation time can be obtained using Matthiessen's[41] rule as $1/\tau_\lambda^p = 1/\tau_\lambda^{pp} + 1/\tau_\lambda^{pe}$, where $1/\tau_\lambda^{pp}$ denotes the *p-p* scattering rate which is related to the three-phonon scattering matrix element[41] and $1/\tau_\lambda^{pe}$ denotes the *p-e* scattering rate which is related to the *e-p* scattering matrix element[35,42].

The *p-p* scattering rate due to three-phonon scattering is given by the Fermi's golden rule (FGR)[43] as

$$\begin{aligned}\frac{1}{\tau_\lambda^{pp}} = \frac{\pi\hbar}{16N}\sum_{\lambda_1\lambda_2}\left|V_{\lambda\lambda_1\lambda_2}\right|^2 &\Big\{\left(n_{\lambda_1}+n_{\lambda_2}+1\right)\delta\left(\omega_\lambda-\omega_{\lambda_1}-\omega_{\lambda_2}\right)+\left(n_{\lambda_1}-n_{\lambda_2}\right) \\ &\times\left[\delta\left(\omega_\lambda+\omega_{\lambda_1}-\omega_{\lambda_2}\right)-\delta\left(\omega_\lambda-\omega_{\lambda_1}+\omega_{\lambda_2}\right)\right]\Big\}\ ,\end{aligned} \tag{2}$$

where $N$ is the total number of phonon modes. $\delta$ is the Dirac delta function, which is approximated by a Gaussian or Lorentzian function[37] in practice. The term $V_{\lambda\lambda_1\lambda_2}$ is the three-phonon scattering matrix element, which is related to the 3rd-order force constants[40].

The *p-e* scattering can be also obtained from FGR[43], under the relaxation time approximation, the scattering rate of phonon mode $\lambda$ is

$$\frac{1}{\tau_\lambda^{pe}} = \frac{2\pi}{\hbar}\sum_{\mathbf{k},i,j}\left|g_{j\mathbf{k}+\mathbf{q},i\mathbf{k}}^{\lambda}\right|^2\left(f_{i\mathbf{k}}-f_{j\mathbf{k}+\mathbf{q}}\right)\times\delta\left(\epsilon_{i\mathbf{k}}-\epsilon_{j\mathbf{k}+\mathbf{q}}+\hbar\omega_\lambda\right), \tag{3}$$

where $g$ is the *e-p* interaction matrix element, $f$ is the Fermi-Dirac distribution function, $\mathbf{k}$ is the electron wave vector, $i$ and $j$ are band indices of electron, $\epsilon$ is the energy of electron, and $\omega$ is the phonon frequency. The *e-p* matrix element which describes an event where an electron at initial state $|i,\mathbf{k}\rangle$ is scattered to $|j,\mathbf{k}+\mathbf{q}\rangle$ by a phonon mode $\lambda = (\mathbf{q},v)$, is defined as[35]

$$g_{j\mathbf{k}+\mathbf{q},i\mathbf{k}}^{\lambda} = \sqrt{\frac{\hbar}{2\omega_\lambda}}\left\langle\psi_{j\mathbf{k}+\mathbf{q}}\left|\partial U_\lambda\right|\psi_{i\mathbf{k}}\right\rangle, \tag{4}$$

where $\psi$ is the ground-state Bloch wave function and $\partial U_\lambda$ denotes the first-order derivative of the Kohn-Sham potential with respect to the phonon displacement. In general, Eq. (3) can be further approximated because of the much smaller energy of phonons than electrons, which is expressed as[27]

$$\frac{1}{\tau_{\lambda}^{pe}} \approx 2\pi \sum_{\mathbf{k},i,j} \left| g_{j\mathbf{k}+\mathbf{q},i\mathbf{k}}^{\lambda} \right|^2 \frac{\partial f(\epsilon_{i\mathbf{k}}, T)}{\partial \epsilon} \times \delta\left(\epsilon_{i\mathbf{k}} - \epsilon_{j\mathbf{k}+\mathbf{q}} + \hbar\omega_{\lambda}\right)\omega_{\lambda}, \tag{5}$$

where $\partial f / \partial \epsilon$ is "Fermi window" that peaks at the Fermi level.

### 2.1.2. Electron thermal conductivity

Combining the BTE and Onsager relations[43], the electron transport properties can be obtained as

$$\sigma_{\alpha\beta} = -\frac{e^2 n_s}{V} \sum_{i\mathbf{k}} \frac{\partial f_{i\mathbf{k}}}{\partial \epsilon} v_{i\mathbf{k}}^{\alpha} v_{i\mathbf{k}}^{\beta} \tau_{i\mathbf{k}}, \tag{6}$$

$$[\sigma S] = -\frac{e n_s}{VT} \sum_{i\mathbf{k}} \left(\epsilon_{i\mathbf{k}} - \mu\right) \frac{\partial f_{i\mathbf{k}}}{\partial \epsilon} v_{i\mathbf{k}}^{\alpha} v_{i\mathbf{k}}^{\beta} \tau_{i\mathbf{k}}, \tag{7}$$

$$K_{\alpha\beta} = -\frac{n_s}{VT} \sum_{i\mathbf{k}} \left(\epsilon_{i\mathbf{k}} - \mu\right)^2 \frac{\partial f_{i\mathbf{k}}}{\partial \epsilon} v_{i\mathbf{k}}^{\alpha} v_{i\mathbf{k}}^{\beta} \tau_{i\mathbf{k}}, \tag{8}$$

where $\sigma_{\alpha\beta}$ is the electrical conductivity and $S_{\alpha\beta}$ is the Seebeck coefficient of 3×3 tensors. $K_{\alpha\beta}$ is related to the electron thermal conductivity $\kappa_e = K - S\sigma ST$, where $T$ is the temperature. The summation in these three equations is over all the electrons enumerated using electronic wave vector $\mathbf{k}$ and band index $i$. The $e$ is the elementary charge, $n_s$ is the number of electrons per state, $V$ is the volume of the primitive cell, $f_{i\mathbf{k}}$ is the Fermi-Dirac distribution, $\epsilon_{i\mathbf{k}}$ is the electron energy, $\mu$ is the chemical potential, $\mathbf{v}_{i\mathbf{k}} = \frac{1}{\hbar}\frac{\partial \epsilon_{i\mathbf{k}}}{\partial \mathbf{k}}$ is the electron velocity, $\alpha$ and $\beta$ denotes the directional components, and $\tau_{i\mathbf{k}}$ is the electron transport relaxation time. The electron transport relaxation time, limited by *e*-*p* scattering, can be obtained by considering the *e*-*p* interactions as[43]

$$\begin{aligned} \frac{1}{\tau_{i\mathbf{k}}} = \frac{2\pi}{\hbar} \sum_{j} \sum_{\lambda} \left| g_{j\mathbf{k}+\mathbf{q},i\mathbf{k}}^{\lambda} \right|^2 &\left\{ \left(n_{\lambda} + f_{j\mathbf{k}+\mathbf{q}}\right) \delta\left(\epsilon_{i\mathbf{k}} + \hbar\omega_{\lambda} - \epsilon_{j\mathbf{k}+\mathbf{q}}\right) \right. \\ &\left. + \left(n_{\lambda} + 1 - f_{j\mathbf{k}+\mathbf{q}}\right) \delta\left(\epsilon_{i\mathbf{k}} - \hbar\omega_{\lambda} - \epsilon_{j\mathbf{k}+\mathbf{q}}\right) \right\}. \end{aligned} \tag{9}$$

### 2.1.3 Analytical models

Actually, the three-phonon scattering strength $V_{\lambda\lambda_1\lambda_2}$ in Eq. (2) is quite nontrivial. In order to obtain an expression of $\kappa_p$, Klemens[4] derived a formula of $V_{\lambda\lambda_1\lambda_2}$ by generalizing the result for long-wavelength phonons to all phonon modes, in which the Debye-like dispersion and ignorance of phonon branch restrictions were assumed. The approximation equation of $V_{\lambda\lambda_1\lambda_2}$ is as following[4,44]

$$\left|V_{\lambda\lambda_1\lambda_2}\right| = B\frac{M\gamma_G}{\sqrt{N}}\frac{\omega_\lambda\omega_{\lambda_1}\omega_{\lambda_2}}{v_g}\sqrt{\frac{\hbar^3}{M^3\omega_\lambda\omega_{\lambda_1}\omega_{\lambda_2}}} , \tag{10}$$

where $B$ is a constant number, $M$ is the total mass of atoms in the unit cell, $\gamma_G$ is the average Grüneisen parameter, and $v_g$ is the phonon group velocity in Debye model. Although this estimation simplifies the complicated term, it is still difficult to calculate the summation in Eq. (2) due to the Dirac delta function. In order to solve this issue, Leibfried[25] used the inverse of the Debye frequency $1/\omega_D$ to approximate the Dirac delta function. With these approximations, the *p-p* scattering rate can by approximated[4,10] as $1/\tau^{pp} = B\gamma_G^2 v_g\left(k_B T/\mu a^3\right)\left(\omega/\omega_D\right)^2$, where $k_B$ is Boltzmann constant and $\mu$ is shear modules. In advance, the formula of phonon thermal conductivity which only considers *p-p* scattering can be written as following[4,10]

$$\kappa_p^{pp} = \frac{3.22}{B\gamma_G^2}\left(\frac{k_B\theta_D}{\hbar}\right)^3\frac{\bar{M}a}{T} , \tag{11}$$

where $\theta_D$ is Debye temperature, $\bar{M}$ is average atomic mass, $a$ is the cube root of unit cell volume and $T$ is temperature. The uncertainties of this model are reflected in an uncertain numerical coefficient of $B$. Leibfried and Schlömann[25] give $B$=0.87, while Klemens[10] gives $B$=2. It should be noted that there exist some debates about the value of $B$. Julian[45] claimed that the value given by Leibfried and Schlömann is smaller by a factor of 0.5 due to a numerical error, which means Julian gave a corrected value of $B$=1.74. Furthermore, Julian[45] also tried to fit the coefficient of $B$ using Grüneisen parameters with the help of digital computers. By using Julian's fitting parameters, Slack[26] presented the expression for the phonon thermal conductivity (only *p-p* scattering) as following

$$\kappa_p^{pp} = \frac{0.849\times 3\sqrt[3]{4}}{20\pi^3\left(1-0.514\gamma_G^{-1}+0.228\gamma_G^{-2}\right)}\times\left(\frac{k_B\theta_D}{\hbar}\right)^2\frac{k_B\bar{M}a}{\hbar\gamma_G^2} . \tag{12}$$

In general, the only consideration of *p-p* scattering of determining the phonon thermal conductivity in metals is not accurate due to the importance of phonon-electron scattering. However, the phonon-electron scattering strength is very complicated as shown in Eq. (3). In order to obtain the expression for the phonon-electron scattering rate corresponding to a phonon relaxation process, the electrons are treated as free electrons of Fermi energy $E_f$ and Fermi velocity $v_0$ to interact with phonons and then the phonon-electron scattering rate can be written as the form[4,10] $1/\tau^{pe} = \pi/3\left(v_g/v_0\right)\left(n_e C_{pe}^2 \omega/\mu a^3 E_f\right)$, where $n_e$ is the number of electrons per atom. $C_{pe}$ is the phonon-electron interaction parameter which has a magnitude comparable to $E_f$ [10,22]. Although the complicated terms have been simplified, but the values in the approximated equations are still difficult to be determined, such as $C_{pe}$. With the help of these approximated relaxation time terms, Klemens[10] gave the phonon thermal conductivity considering both *p-p* and *p-e* scattering schemes as following

$$\kappa_p^{pp+pe} = \kappa_p^{pp}\left[1 - \frac{\omega_i}{\omega_D}\ln\left(\frac{\omega_D}{\omega_i}+1\right)\right], \tag{13}$$

where $\omega_i = \omega_D\left(v_g/v_0\right)\left(n_e\pi/3B\gamma_G^2\right)\left(C_{pe}^2/k_B T E_f\right)$. With further simplification, we find that $\omega_i/\omega_D$ equals $\tau^{pp}/\tau^{pe}$ with the above approximated relaxation time terms of $\tau^{pp}$ and $\tau^{pe}$. It should be noted that the phonon thermal conductivity calculated by using these analytical formulas will be compared with the first-principles calculations, which will help us to understand the accuracy of these approximated models.

### 2.2. First-principles calculations

The first-principles calculations including density functional theory (DFT) and density functional perturbation theory (DFPT) are carried out using Quantum Espresso package[46] to predict the phonon and electron thermal transport in these metals by considering *p-p* and *p-e* scatterings. In *p-p* scattering rate calculations, the second-order interatomic force constants (2$^{nd}$-IFCs) are obtained using DFPT and the 3$^{rd}$-IFCs are obtained using the finite-difference supercell methods in which the forces are extracted from the self-consistent field calculation of displaced supercell configurations. In order to calculate the 3$^{rd}$-IFCs, the supercell is created by using thirdorder.py package[47]. The size of supercell and the nearest neighbors are provided in the supporting information of Sec. S2. In the *p-e* scattering rate calculations, the phonon perturbation is first calculated using DFPT as implemented in Quantum Espresso[46] and then the *e-p* scattering matrix element is calculated in Electron-Phonon Wannier (EPW) package[48]. The *e-p* scattering matrix element is initially obtained on coarse electron and phonon wave vector grids and then interpolated to denser electron and phonon wave vector grids using the maximally localized Wannier functions[49] basis as implemented in EPW[48]. The denser meshes of wave vector for calculating the *e-p* scattering rate are listed in Sec. S3. In these calculations, the norm-conserving pseudopotentials[50] are used. The exchange and correlation (XC) functional is treated by local density approximation (LDA)[51] or Generalized Gradient Approximation (GGA)[52] in our calculations. The choice of XC functional depends on the material, and it is determined by searching the literature with suggested XC functional for the corresponding material.

Importantly, it should be noted that the following factors may have effects on the predicted values of thermal conductivity and electrical conductivity from first-principles calculations: (1) pseudopotentials used in DFT calculations[27], (2) the number of **k** and **q**-mesh used in the interpolation process during electron-phonon scattering rate calculations[37], (3) the method such as the relaxation time approximation (RTA) or iterative scheme used for calculating the thermal conductivity[53]. Here, we try to choose the pseudopotentials which make sure the DFT calculated electrical conductivity matches well with the experimental values. Also, we choose the moderate **k** and **q**-mesh in electron-phonon scattering rate calculations based on the balance of accuracy and computational cost (see Sec. S3). In addition, we use the RTA method which can guarantee the denser **q**-mesh used in three-phonon scattering rate calculations, and it has been reported that the difference of the calculated thermal conductivity between RTA and iterative method is almost negligible for metals[53]. Actually, it is not easy to absolutely resolve all these effects into the calculations, but we have tried our best to make sure our calculations are available and believable.

## 3. Results and discussion

### 3.1 Electrical conductivity, electron thermal conductivity and phonon thermal conductivity

Table 1. The DFT predictions of electrical conductivity $\sigma$ and total thermal conductivity $\kappa_{total}^{\mathrm{DFT}} = \kappa_e + \kappa_p$ are compared to experimental values at room-temperature[10,22,54–58]. $\kappa_e$ denotes for electron thermal conductivity, $\kappa_p$ for phonon thermal conductivity, and $\kappa_{total}^{\mathrm{Exp.}}$ for experimental

value. $L_0 = \pi^2 k_B^2 / (3e^2) = 2.44\times10^{-8}\,\mathrm{V}^2\cdot\mathrm{K}^{-2}$ is the Sommerfeld value of the Lorenz ratio[10]. $L = \kappa_e/(\sigma T)$, where $T$ is temperature.

| Material | $\sigma$ ($\times10^7\ \Omega^{-1}\mathrm{m}^{-1}$) | | $\kappa_p$ (W/mK) | $\kappa_e$ (W/mK) | $\kappa_{total}^{\mathrm{DFT}}$ (W/mK) | $\kappa_{total}^{\mathrm{Exp.}}$ (W/mK) | $\frac{\kappa_p}{\kappa_{total}^{\mathrm{DFT}}}$ | $\frac{L}{L_0}$ |
|---|---|---|---|---|---|---|---|---|
| | DFT | Exp. | DFT | DFT | DFT | Exp. | DFT, % | DFT |
| Ag | 6.37 | 6.21 | 5.59 | 475.80 | 481.39 | 436.00 | 1.16 | 1.02 |
| Au | 3.50 | 4.50 | 2.72 | 264.94 | 267.66 | 318.00 | 1.02 | 1.03 |
| Cu | 5.17 | 5.78 | 17.61 | 360.90 | 378.52 | 402.00 | 4.65 | 0.96 |
| Al | 3.41 | 4.12 | 8.14 | 225.76 | 233.89 | 237.00 | 3.48 | 0.91 |
| Mg | 2.48 | 2.30 | 7.19 | 178.20 | 185.39 | 153.00 | 3.88 | 0.98 |
| Pt | 1.32 | 1.02 | 6.28 | 86.71 | 92.99 | 71.90 | 6.75 | 0.90 |
| Pd | 0.90 | 1.03 | 12.99 | 68.71 | 81.70 | 71.70 | 15.90 | 1.05 |
| Ni | 1.43 | 1.60 | 14.50 | 101.09 | 115.59 | 93.00 | 12.54 | 0.97 |
| Ti | 0.21 | 0.25 | 6.68 | 28.61 | 35.29 | 22.30 | 18.93 | 0.93 |
| Co | 1.15 | 1.67 | 12.20 | 70.72 | 82.92 | 99.00 | 14.71 | 0.84 |
| Mn | 0.08 | 0.07 | 3.28 | 5.05 | 8.33 | 7.80 | 39.38 | 0.84 |
| NiAl | 1.00 | 1.02 | 6.33 | 69.29 | 75.62 | 76.00 | 8.37 | 0.94 |
| $Ni_3Al$ | 0.38 | 0.30 | 5.22 | 29.72 | 34.94 | 28.50 | 14.94 | 1.07 |
| TiAl | 0.11 | 0.13 | 4.57 | 7.80 | 12.37 | 11.50 | 36.94 | 0.95 |
| FeAl | 0.13 | 0.18 | 3.80 | 8.83 | 12.63 | 12.00 | 30.09 | 0.92 |
| CoAl | 0.55 | 0.71 | 4.83 | 44.48 | 49.31 | 37.00 | 9.80 | 1.10 |
| $Cu_3Au$ | 1.74 | 1.85 | 1.89 | 126.27 | 128.15 | 157.20 | 1.47 | 0.99 |
| CuAu | 1.75 | 1.32 | 2.85 | 126.17 | 129.02 | 167.00 | 2.21 | 0.99 |

By implementing the first-principles calculations, the electrical conductivity $\sigma$, phonon $\kappa_p$, and electron $\kappa_e$ thermal conductivity are obtained as shown in Table 1. First, we can see that the predicted $\sigma$ in general agree well with experimental data[10,22,54–58]. The difference is in the range of 2% to 31%, and most of them are within 15%, which is acceptable. Since the electrical conductivity is related to band structure, the electron velocity, individual electron-phonon scattering matrix element and the Brillouin zone integration details, the good agreement of calculated $\sigma$ with experimental results indicates that band structure calculation and *e-p* coupling prediction are reliable. On the other hand, we can also compare the DFT predictions of total thermal conductivity $\kappa_{total}^{\mathrm{DFT}} = \kappa_e + \kappa_p$ at room-temperature with experimental values[10,22,54–58] as presented in Table 1. It can be seen that the DFT predicted $\kappa_{total}^{\mathrm{DFT}}$ also agree well with experimental values $\kappa_{total}^{\mathrm{Exp.}}$. These comparisons and validations provide confidence for us to further analyze *e-p* coupling strength and its effect on phonon thermal conductivity.

From the values of $\kappa_p$ and $\kappa_e$ in Table 1, we note that the phonon thermal conductivity ranges from 2 to 18 W/mK. The ratio of phonon thermal conductivity to total thermal conductivity $\kappa_p / \kappa_{total}$ can be smaller than 2% or as large as 40% at 300 K. 8 out of 18 metals have phonon contributions of more than 10%. Therefore, the contribution of the phonon to the total thermal conductivity in metals cannot be neglected, at least not for all metals. Furthermore, $\kappa_p$ can play a more important role in the thermal conductivity of metal nanostructure due to the significant reduction of $\kappa_e$ at nanostructure[29]. Therefore, our calculation results show the necessity of first-principles investigation on the phonon thermal transport in metals.

In addition, the predicted $\kappa_p^{pp+pe}$ from first-principles are compared with the predictions by Klemens model[10] (Eq. (13)), and the comparisons are shown in Fig. 1. It should be noted that the average relaxation time of $\tau^{pp}$ and $\tau^{pe}$ calculated from first-principles were used in Eq. (13). We can see that the Pearson correlation between the first-principles prediction and the theoretical Klemens prediction is 0.25. It indicates that the Klemens model fails to accurately predict the phonon thermal conductivity. This is not surprising since the Klemens model was derived based on the assumption of Debye approximation, free electrons interacting with phonons, and the long wavelength phonons[10]. Therefore, the previous phonon thermal conductivity estimations of metals from Klemens model have large uncertainty and must be used with care.

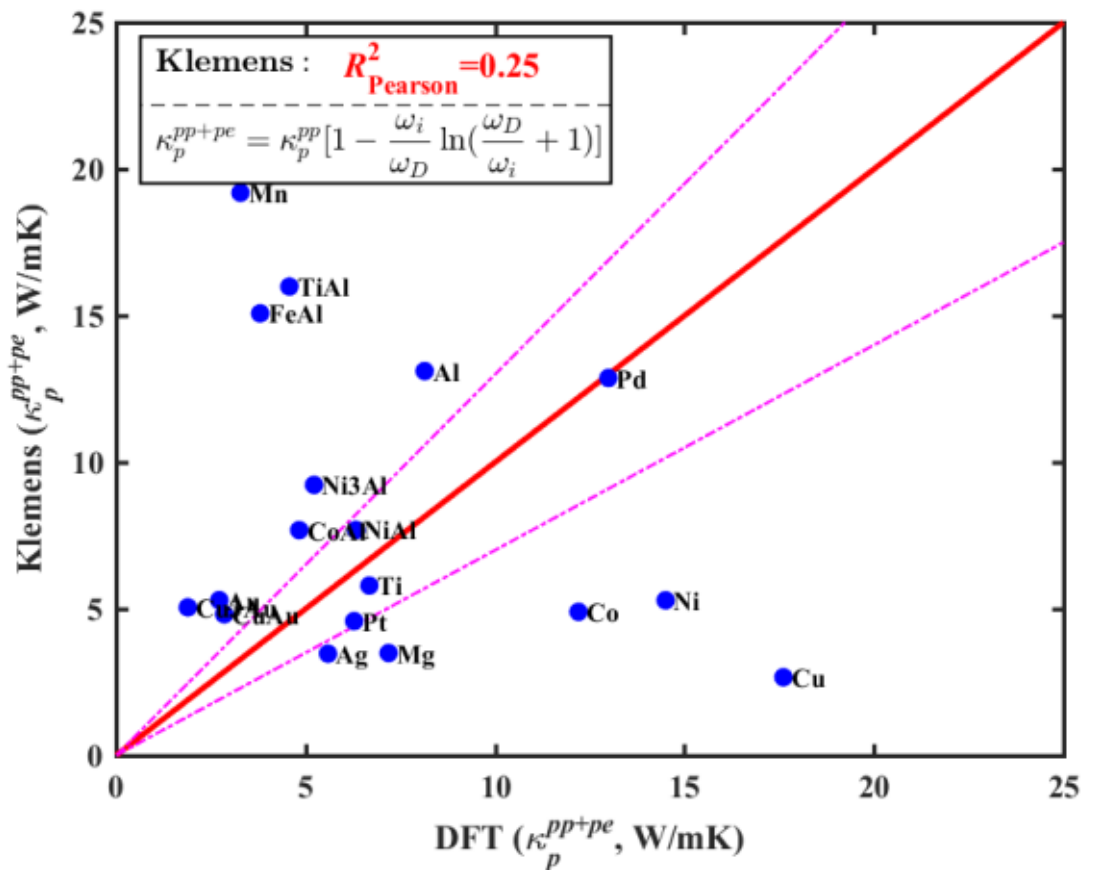

Fig. 1. Phonon thermal conductivity $\kappa_p^{pp+pe}$ predictions at a temperature of 300 K from first-principles calculations and Klemens model[10], which include both *p*-*p* and *p*-*e* scattering. The dash line represent ± 30% error.

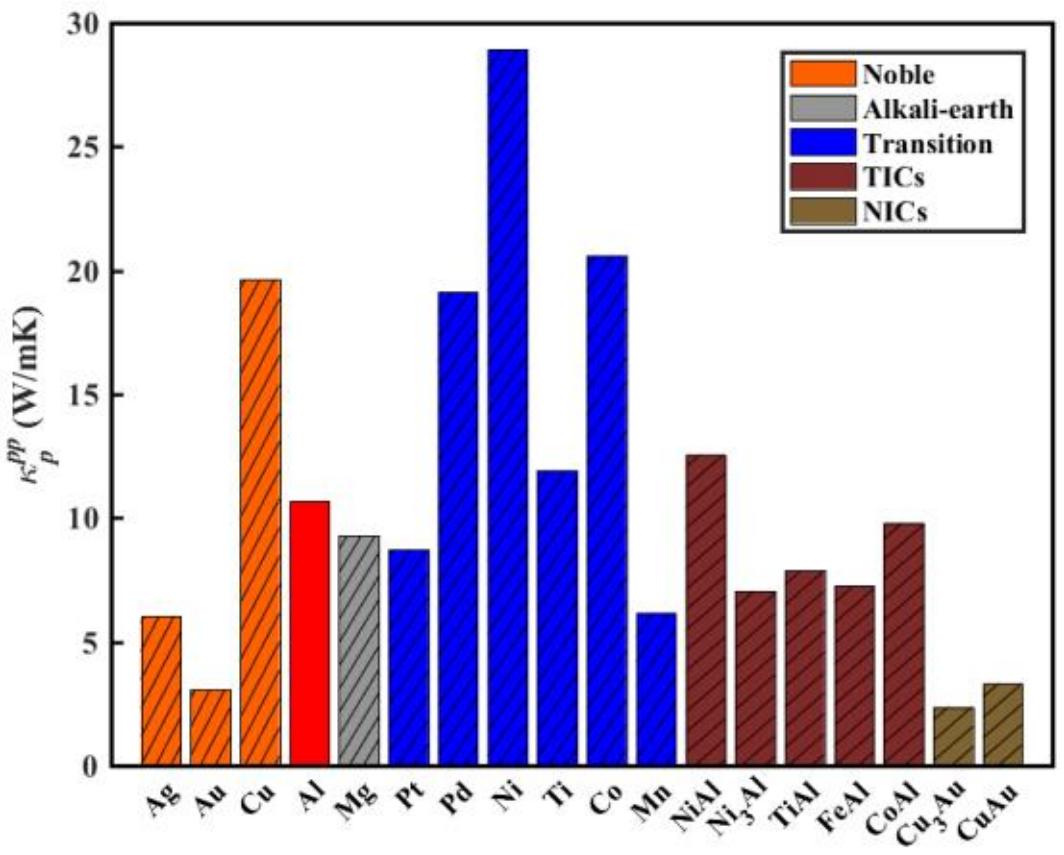

Fig. 2. Phonon thermal conductivity $\kappa_p^{pp}$ of noble, alkali-earth, transition, TICs and NICs at 300 K. Here, only *p*-*p* scattering is considered in the calculation of phonon thermal conductivity.

### 3.2 Phonon thermal conductivity with only *p-p* scattering

The above calculations provide relatively reliable data for the phonon thermal conductivity of different types of metals. To gain further insights into the phonon thermal conductivity of metals, further analysis is necessary. As shown in Eqs. (2) and (3), the phonon thermal conductivity in metals is affected by both *p*-*p* and *p*-*e* scattering. Therefore, we firstly investigate the phonon thermal conductivity with only considering *p*-*p* scattering, and then further consider electron-phonon coupling effect on the phonon thermal conductivity.

The first-principles calculated phonon thermal conductivity $\kappa_p^{pp}$ (only *p*-*p* scattering is considered) at 300 K are shown in Fig. 2. The values of $\kappa_p^{pp}$ are within the range of 2 - 30 W/mK. While most of the metals have $\kappa_p^{pp}$ smaller than or approach to 10 W/mK, there are a few exceptions, including Cu, Pd, Ni, and Co. These are all elemental metals with relatively small atomic masses. Except for Cu which is a noble metal, all others are transition metals. It is well known that the phonon thermal conductivity is related to atomic mass, bonding strength, and anharmonicity of lattice. It is not surprising that these materials (like Pd, Pt, Ti, Ni, Co, Mn) have relatively higher $\kappa_p^{pp}$ than that of noble metals (like Ag, Au), since transition metals generally have stronger bonding[59] as compared to other metals. The binding energy of these materials are shown in Table 2, which supports this statement.

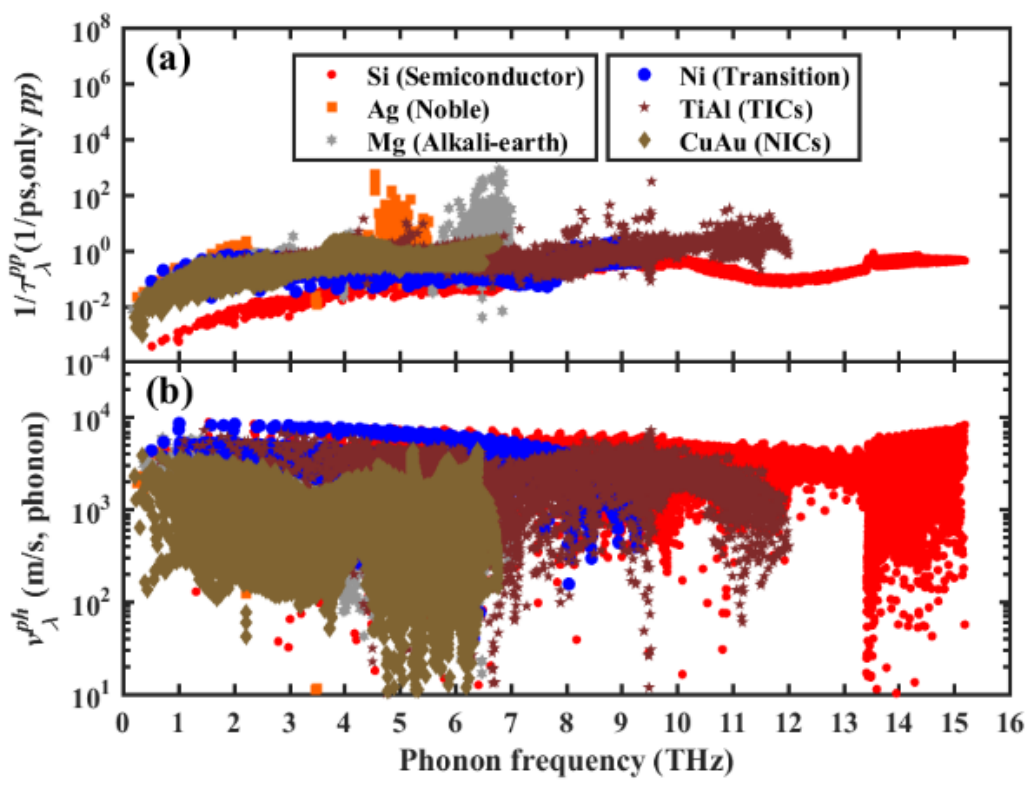

Fig. 3. (a) Phonon-phonon scattering rate and (b) phonon group velocity for Ag (noble), Mg (alkali-earth), Ni (transition), TiAl (TICs), CuAu (NICs) and Si (semiconductor) at 300 K.

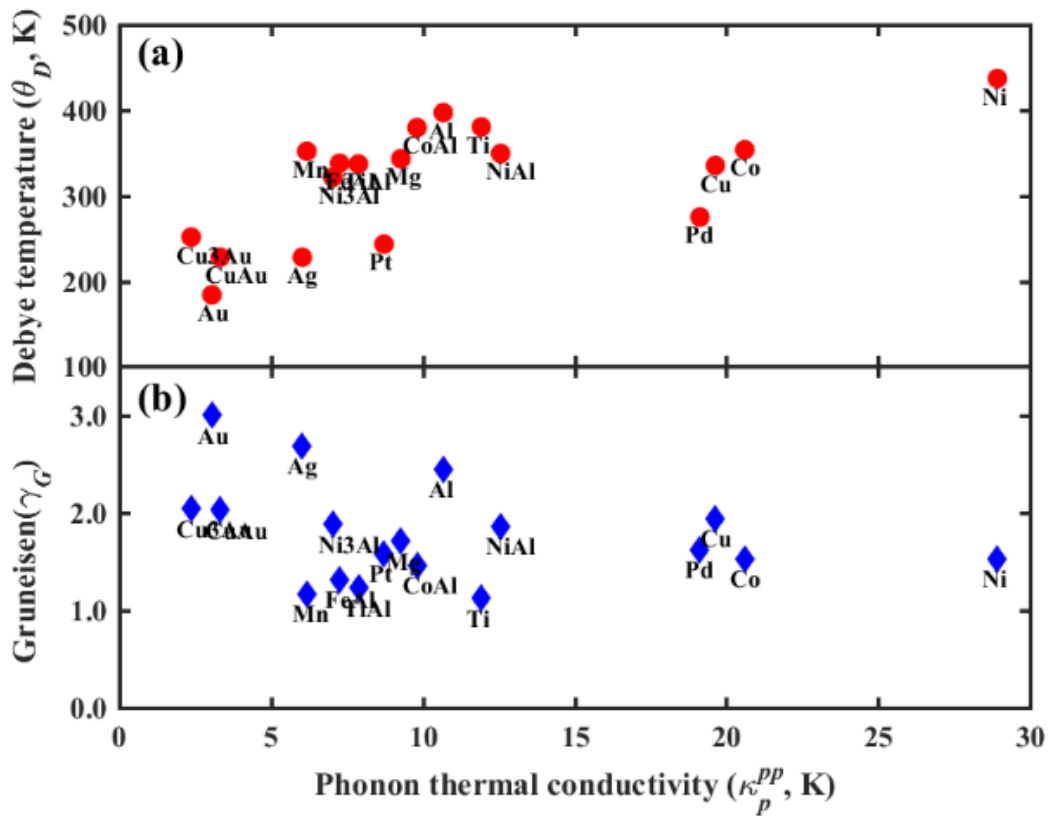

Fig. 4. The phonon thermal conductivity $\kappa_p^{pp}$ variation with Debye temperature and Grüneisen parameter $\gamma_G$ at 300 K.

From Eq. (1), one can see that the phonon thermal conductivity is related to both group velocity and relaxation time. In order to make further analysis on the phonon thermal conductivity, we plot the mode-dependent *p*-*p* scattering rate $1/\tau_\lambda^{pp}$ and phonon group velocity $v_\lambda^{ph}$ for a few representative materials for comparison, including Ag (noble), Mg (alkali-earth), Ni (transition), TiAl (TICs), and CuAu (NICs) , as shown in Fig. 3. We can see that the *p*-*p* scattering rates for these metals are comparable from Fig. 3(a). However, the phonon group velocity of transition metal (Ni) is larger than that of noble metal (Ag) and it is also larger for TICs (TiAl) compared to NICs (CuAu) as shown in Fig. 3(b). From the comparison, we can see that the phonon group velocity is the dominant factor in causing the differences in phonon thermal conductivity. In addition, by comparing the phonon group velocity and *p*-*p* scattering rate of the metals with that of good semiconductor of silicon (~150 W/mK) as shown in Fig. 3, we find that the metals have much larger *p*-*p* scattering rate and lower phonon group velocity than silicon. In fact, this is because covalent bonding in silicon is usually stronger than the metallic bonding in metals[59].

Furthermore, in the analytical models, the strength of *p*-*p* scattering and phonon group velocity are usually quantified by the Grüneisen parameter $\gamma_G$ and Debye temperature $\theta_D$ , respectively. Here, we also present the predicted values of $\gamma_G$ and $\theta_D$, as shown in Table 2. It should be mentioned that we compared the first-principles predictions of $\gamma_G$ and $\theta_D$ with experimental values and they agree well (see Sec. S4). The variation of $\kappa_p^{pp}$ with $\gamma_G$ and $\theta_D$ are also plotted as shown in Fig. 4. Overall, we can see that $\kappa_p^{pp}$ is larger with higher $\theta_D$ and smaller $\gamma_G$. This is also consistent with the general theory for phonon thermal conduction, which says that the larger phonon group velocity (strong bonding with high $\theta_D$) and smaller *p-p* scattering rate (weak anharmonicity with small $\gamma_G$) result in larger phonon thermal conductivity. The Pearson correlation coefficients between $\kappa_p^{pp}$ and the parameters are 0.26 and 0.12 for $\theta_D$ and $\gamma_G$, respectively. This further indicates that the group velocity is more important in determining the phonon thermal conductivity, but neither of the two parameters can be directly used to evaluate $\kappa_p^{pp}$.

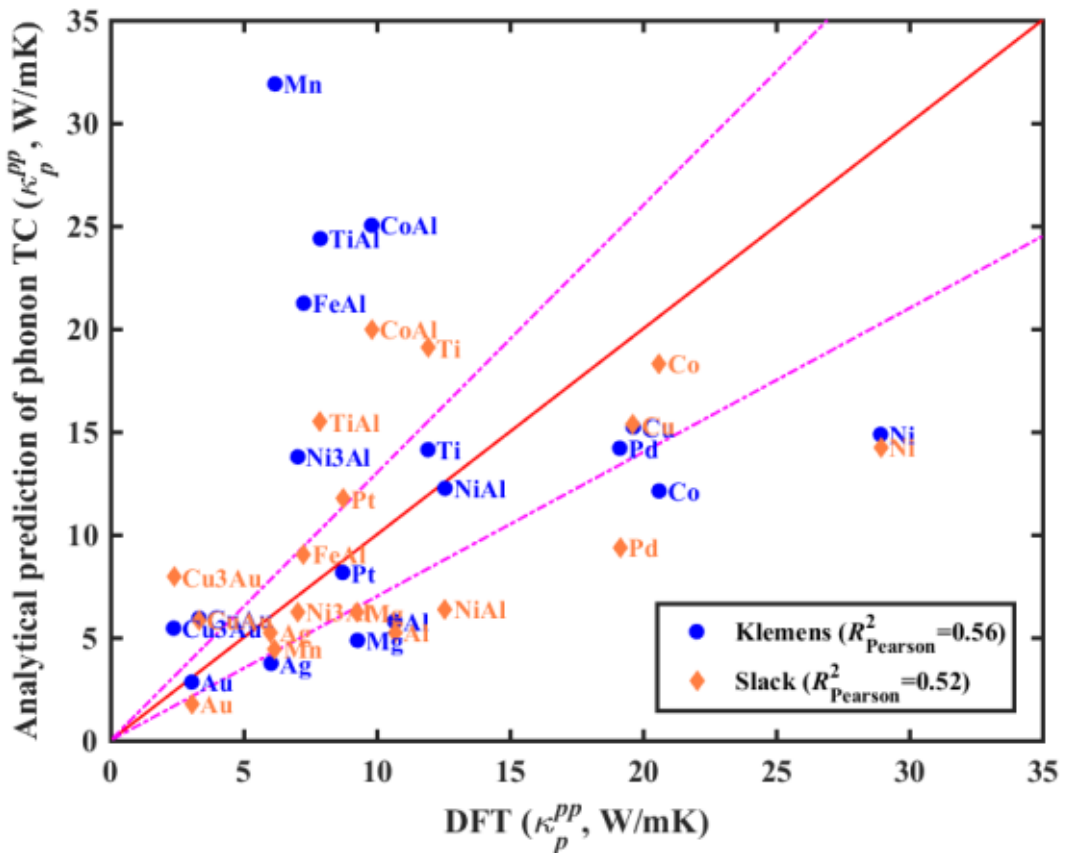

Fig. 5. Phonon thermal conductivity predictions at a temperature of 300 K from first-principles calculations and theoretical calculations using Klemens[10] and Slack model[26], which only consider the *p-p* scattering. The dash line represent ± 30% error.

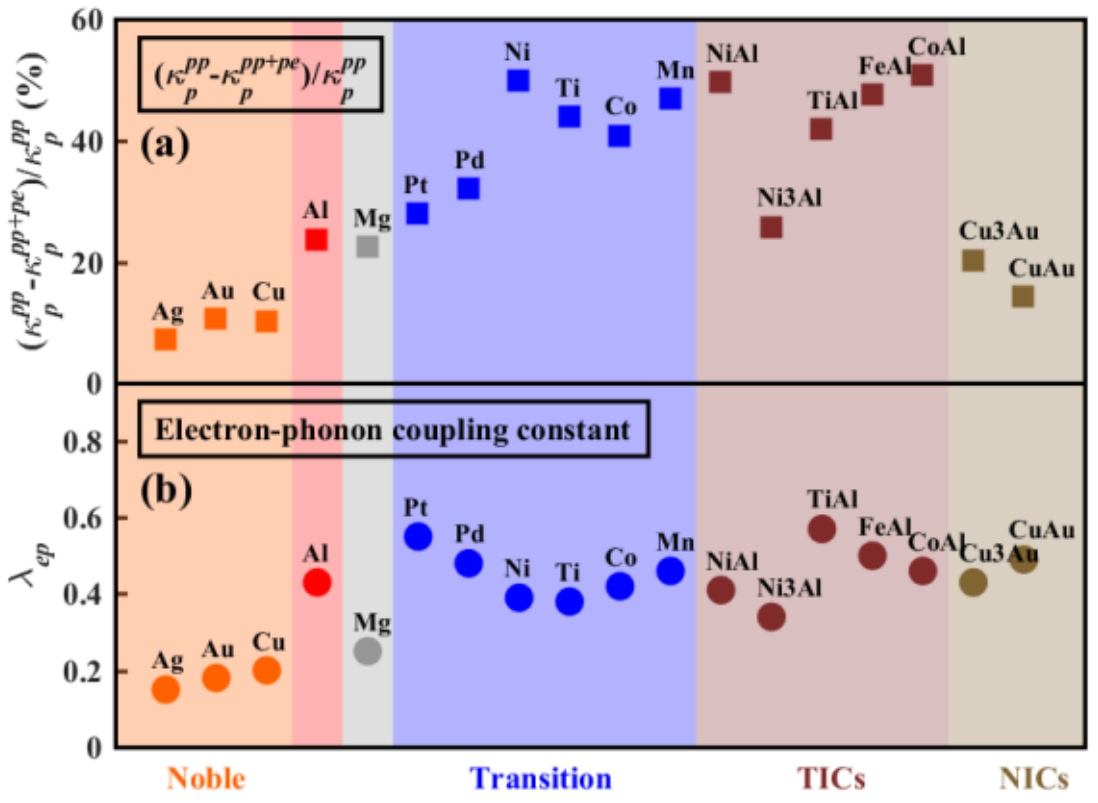

Fig. 6. (a) Percentage of the reduction of phonon thermal conductivity $\kappa_p^{pp}$ induced by electron-phonon coupling effect. (b) Electron-phonon coupling constant.

Since classical thermal conductivity models were widely used to estimate the phonon thermal conductivity of metals[4,60], we can also check their accuracy. For this purpose, the predicted $\kappa_p^{pp}$ from first-principles are compared with the predictions by the widely used Klemens model[4] (Eq. (11)) and Slack model[26] ( Eq. (12)), and the comparisons are shown in Fig. 5. It should be noted that the first-principles calculated Debye temperature and Grüneisen parameter are used in both Klemens and Slack model. We can see that the Pearson correlation between first-principles prediction and theoretical prediction is only 0.56 and 0.52 for Klemens and Slack model, respectively. It indicates that these analytical models fail to accurately predict the phonon thermal conductivity. This is not surprising since these analytical models generally adopt the Debye approximation and the long wavelength assumption is employed in Klemens model[10]. Comparing to semiconductors and dielectrics[44], these models are not reliable for metals, presumably because these models were originally developed for non-metallic materials[25]. As such, the previous phonon thermal conductivity estimations from these models[10,29,61] have large uncertainty and must be used with care.

### 3.3 Electron-phonon coupling effect on phonon thermal conductivity

Actually, the *p-e* scattering is an important scattering term in phonon scattering process for metals and it should be rigorously considered. It was believed that the *p-e* scattering term has relatively small contribution to phonon thermal conductivity at medium temperature range[10,22,27,37]. However, such statement cannot be completely supported by our results. Here, we quantified the reduction $(\kappa_p^{pp} - \kappa_p^{pp+pe}) / \kappa_p^{pp}$ of $\kappa_p^{pp}$ after including the *p-e* scattering effect as shown in Fig. 6 and the data values are also presented in Table 2. It can be seen that the *e-p* coupling effect on phonon thermal conductivity $\kappa_p^{pp}$ varies strongly with different metals.

Table 2. DFT predicted values of Debye temperature $\theta_D$ , Grüneisen parameter $\gamma_G$ , $\kappa_p^{pp}$ (only *p-p* scattering is considered) and $\kappa_p^{pp+pe}$ (both *p-p* and *p-e* scattering are considered), and reference data[27,29,37] for $\kappa_p^{pp+pe}$ . DFT predicted *e-p* coupling constant $\lambda_{ep}$ and experimental values[62–65] of $\lambda_{ep}$ .

| Material | Debye (K) | Grüneisen parameter | Binding energy | Phonon thermal conductivity at 300 K (W/mK) | | | *e-p* coupling constant | |
|---|---|---|---|---|---|---|---|---|
| | $\theta_D$ (DFT) | $\gamma_G$ (DFT) | - (DFT) | $\kappa_p^{pp}$ (DFT) | $\kappa_p^{pp+pe}$ (DFT) | $\kappa_p^{pp+pe}$ (Literature) | $\lambda_{ep}$ (DFT) | $\lambda_{ep}$ (Exp.[d]) |
| Ag | 228.18 | 2.69 | 2.61 | 6.03 | 5.59 | 9.3[a], 5.2[b], 4.0[c] | 0.15 | 0.13 |
| Au | 184.17 | 3.00 | 3.38 | 3.05 | 2.72 | 5.0[a], 2.6[b], 2.0[c] | 0.18 | 0.15 |
| Cu | 335.4 | 1.94 | 4.85 | 19.64 | 17.61 | 22.2[a], 16.9[b] | 0.20 | 0.14 |
| Al | 397.26 | 2.45 | 3.53 | 10.67 | 8.14 | 21.1[a], 5.8[b], 6.0[c] | 0.43 | 0.43 |
| Mg | 343.35 | 1.71 | 4.82 | 9.27 | 7.19 | - | 0.25 | 0.27 |
| Pt | 243.48 | 1.58 | 5.70 | 8.72 | 6.28 | 8.3[a], 5.8[b] | 0.55 | 0.66 |
| Pd | 275.13 | 1.62 | 5.18 | 19.13 | 12.99 | - | 0.48 | 0.69 |
| Ni | 437.16 | 1.53 | 5.21 | 28.92 | 14.50 | 42.2[a], 23.2[b] | 0.39 | 0.31 |
| Ti | 380.54 | 1.12 | 5.18 | 11.92 | 6.68 | - | 0.38 | 0.38 |
| Co | 353.83 | 1.53 | 5.99 | 20.60 | 12.20 | - | 0.42 | - |
| Mn | 352.09 | 1.17 | 8.14 | 6.17 | 3.28 | - | 0.46 | - |
| NiAl | 349.45 | 1.86 | - | 12.56 | 6.33 | - | 0.41 | - |
| $Ni_3Al$ | 321.72 | 1.88 | - | 7.03 | 5.22 | - | 0.35 | - |
| TiAl | 337.24 | 1.23 | - | 7.88 | 4.57 | - | 0.57 | - |
| FeAl | 338.01 | 1.32 | - | 7.25 | 3.80 | - | 0.50 | - |
| CoAl | 379.66 | 1.46 | - | 9.81 | 4.83 | - | 0.46 | - |
| $Cu_3Au$ | 251.71 | 2.04 | - | 2.37 | 1.89 | - | 0.43 | - |
| CuAu | 228.17 | 2.03 | - | 3.32 | 2.85 | - | 0.49 | - |

[a] is referred to[29].

[b] is referred to[27].

[c] is referred to[37].

[d] is referred to[62–65].

To understand how the scattering with electrons affects the phonon thermal conductivity, we first examine the expression of *p*-*e* scattering rate $1/\tau_\lambda^{pe}$ in Eq. (5) to figure out the determining factors on the strength of *p*-*e* scattering rate. With the summation of the product of $\left|g_{j\mathbf{k+q},i\mathbf{k}}^{\lambda}\right|^2$ and $\frac{\partial f(\epsilon_{i\mathbf{k}},T)}{\partial\epsilon}\delta\left(\epsilon_{i\mathbf{k}}-\epsilon_{j\mathbf{k+q}}+\hbar\omega_\lambda\right)\omega_\lambda$ in Eq. (5), we can see that stronger *e*-*p* coupling matrix element ($\left|g_{j\mathbf{k+q},i\mathbf{k}}^{\lambda}\right|^2$) and higher phonon frequency ($\omega_\lambda$) induce larger *p*-*e* scattering rate if there are enough electron states within the Fermi window ($\frac{\partial f(\epsilon_{i\mathbf{k}},T)}{\partial\epsilon}$). Therefore, we put forward that the strong *p*-*e* scattering rate generally follows the three conditions: (I) enough electron states around the Fermi surface, (II) high phonon frequency, (III) large *e*-*p* coupling matrix element. The high electron density of states within the Fermi window provides more available electron state for *e*-*p* scattering, which results in a stronger *e*-*p* interaction. The strong *e*-*p* interaction is manifested by a high *e*-*p* coupling constant $\lambda_{ep}$ [33,35] which describes all the possible combinations with $\epsilon_{i\mathbf{k}}$ and $\epsilon_{j\mathbf{k+q}}$ on the Fermi surface under the perturbation of phonon with frequency $\omega_\lambda$. In other words, the satisfaction of condition I and III makes high $\lambda_{ep}$, but it does not always ensure large *p*-*e* scattering rate. This is because the phonon energy ( $\hbar\omega_\lambda$ ) is much smaller than electron energy ( $\epsilon_{i\mathbf{k}}$ ) for most materials, which means that condition II is hard to be satisfied to ensure a large value of $\delta\left(\epsilon_{i\mathbf{k}} - \epsilon_{j\mathbf{k+q}} + \hbar\omega_\lambda\right)\omega_\lambda$ in Eq. (5). Therefore, if the material have high phonon frequencies which equal the energy difference between $\epsilon_{j\mathbf{k+q}}$ and $\epsilon_{i\mathbf{k}}$ states, the $\delta\left(\epsilon_{i\mathbf{k}} - \epsilon_{j\mathbf{k+q}} + \hbar\omega_\lambda\right)\omega_\lambda$ will be larger.

Based on these theoretical understanding and combining the data in Fig. 6 and Table 2, we can draw the following conclusions. Firstly, transition metals (Pt, Pd, Ni, Ti, Co and Mn) have a stronger *e-p* coupling effect than that of noble metals (Au, Ag, Cu). This is because the transition metals have higher electron density of states near Fermi surface (see the electron density of states in Sec. S5), larger *e-p* coupling constant $\lambda_{ep}$ and higher phonon frequency (see the phonon dispersion curve in Sec. S5) compared to that of noble metals, which is consistent with the three conditions for strong *e-p* coupling effect. Secondly, if the material satisfies only one of the conditions (high $\lambda_{ep}$ or high $\omega_\lambda$ ), the *e-p* coupling effect are weaker than that of materials with both high $\lambda_{ep}$ and high $\omega_\lambda$ . For example, the TICs (FeAl, CoAl, NiAl, $Ni_3Al$ and TiAl) have both larger $\lambda_{ep}$ (> 0.2) and higher phonon frequency $\omega_\lambda$ compared to NICs (CuAu and $Cu_3Au$) with only higher $\lambda_{ep}$ but lower $\omega_\lambda$ . In other words, TICs atoms are lighter than NICs and will cause higher phonon frequency and be easier to satisfy condition II under the condition of comparable $\lambda_{ep}$ . Therefore, TICs have a stronger *e-p* coupling effect. Thirdly, the CuAu and $Cu_3Au$ have stronger *e-p* coupling effect than that of Cu and Au, which is due to the participation of optical phonons in the electron-phonon interactions within CuAu and $Cu_3Au$ but no optical phonons in Cu and Au. This can be explained through the Eliashberg spectral function $\alpha^2F(\omega)$ which is generally used to quantify the phonon frequency contribution to the *e-p* coupling strength[33]. The $\alpha^2F(\omega)$ of CuAu and $Cu_3Au$ is plotted as shown in Fig. 7, we can see that the optical phonon contribution to $\alpha^2F(\omega)$ is considerable and it cannot be neglected. This result tells us that the optical phonon in CuAu and $Cu_3Au$ makes great contribution to the *e-p* coupling. Additionally, the electron density of states of Al and alkali-earth metal (Mg) behave like the density of states of free electron, which induces high electron density of states within the Fermi window (see Sec. S5). Therefore, the *e-p* coupling effects in Al and Mg are stronger compared to noble metals which have low electron density of states within Fermi window although they have comparable $\omega_\lambda$ (see Sec. S5)

### 3.4 Electron and phonon contribution to thermal conductivity

It is well known that the electrical conductivity and total thermal conductivity can be measured directly in experiments. However, it is hard to separately measure the electron and phonon thermal conductivity directly. In general, the Wiedemann-Franz law is widely used to evaluate the electron thermal conductivity, in which the electron thermal conductivity is determined through the equation $\kappa_e = L\sigma T$, where $\kappa_e$ is electron thermal conductivity, $L$ is Lorenz number, $\sigma$ is electrical conductivity and $T$ is temperature. Furthermore, the phonon thermal conductivity component can then be obtained by $\kappa_{ph} = \kappa_{tot} - L\sigma T$. For simplicity, people usually use the Sommerfeld value[18] with constant $L_0 = 2.44\times10^{-8}\,\mathrm{V}^2\cdot\mathrm{K}^{-2}$ to evaluate the electron thermal conductivity. However, the $L = L_0$ is only valid at low temperature ($T << \theta_D$, electron-impurity elastic scattering dominant) or high-temperature region ($T >> \theta_D$, electron-phonon elastic scattering dominant), and $L$ will deviate from $L_0$ at intermediate temperature region due to the inelastic *e-p* scattering[22]. Here, by using the first-principles predicted $\kappa_e$ and $\sigma$, we calculated the $L$ at 300 K and compared it with the $L_0$ as shown in Table 1. It can be seen that the deviation between $L$ and $L_0$ is within 1% - 17%, which indicates that the general treatment of evaluating electron thermal conductivity by using $L_0$ should be careful even for metals.

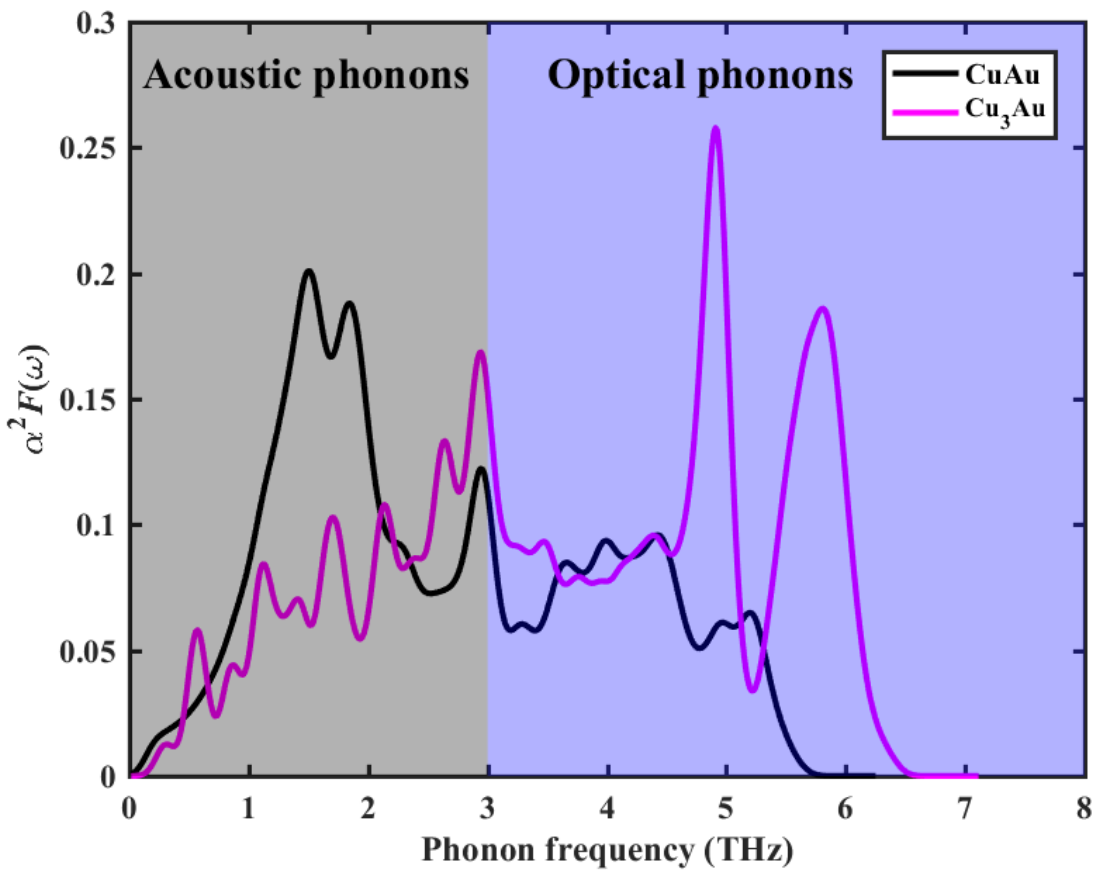

Fig. 7. Variation of Eliashberg spectral function $\alpha^2 F(\omega)$ with phonon frequency for CuAu and $Cu_3Au$. The gray region represents the phonon frequency for acoustic phonons and the blue region for optical phonons. It should be claimed that the separating point of the acoustic and optical phonon is about at 3THz for both CuAu and $Cu_3Au$, which is shown in Sec. S5.

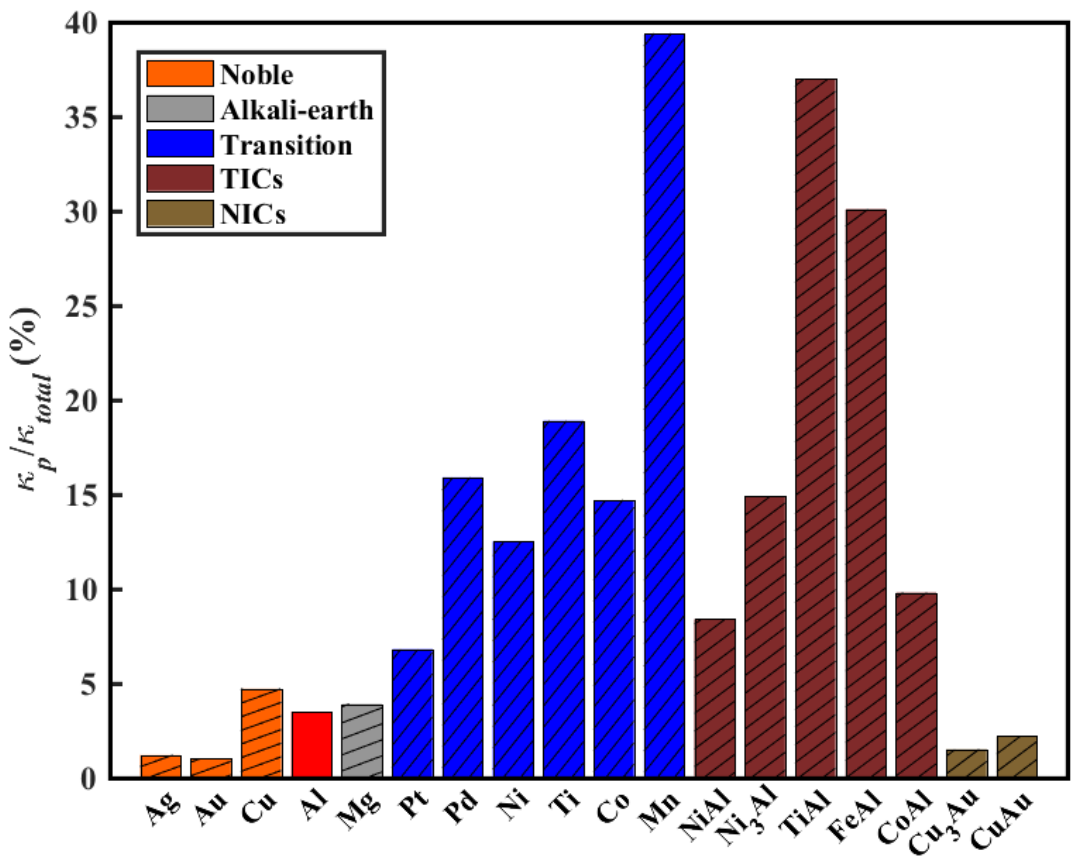

Fig. 8. The percentage of phonon thermal conductivity contributing to total thermal conductivity $\kappa_p / \kappa_{total}$ for these 18 metals.

In addition, we also analyzed phonon contribution to the thermal conduction through the ratio of $\kappa_p$ to $\kappa_{total}$ as shown in Fig. 8. We can see that the phonon contribution can be neglected with $\kappa_p / \kappa_{total}$ less than 10% for noble metals, alkali-earth and NICs, but it could be non-negligible for transition and TICs with $\kappa_p / \kappa_{total}$ ranging between 10% and 40% though their absolute values of $\kappa_p$ range in 3 - 15 W/mK. Actually, for noble metals, the *e-p* scattering in electron transport is very weak compared to transition metals, which makes the electrical conductivity is much larger than that of transition metals as seen in Table 1. Therefore, the electron thermal conductivity is high for noble metals and it gives much smaller phonon component of thermal conductivity. On the other hand, the phonon component of thermal conductivity in the transition metals and intermetallic compounds is comparable as discussed in Section 3.3. Combing these two factors in transition metals and intermetallic compounds, it is not surprising that the total thermal conductivity of them is low as seen in Table 1 and the phonon component of thermal conductivity is large as seen in Fig. 8 compared to the noble metals.

### 3.5 Electron and phonon mean free path

With the development of nanoelectronic devices[13–17], the metal structures with nanoscale dimension were widely used. However, the thermal conductivities of nanostructures are significantly different from their bulk values, and the size effect generally induces reductions due to scattering of electrons and phonons at surfaces and by grain boundaries. In order to figure out the size effect on the thermal conductivity, the mean free path (MFP) for both phonons and electrons were calculated. The MFP denoted by Λ is a measure of the distance traveled by a carrier between scattering events and is the product of the magnitude of its velocity and lifetime (e.g., for phonon mode λ, $\Lambda_\lambda = |v_\lambda| \tau_\lambda$).

The average mean free path can have different definitions. Also, the mean free path for electric conduction and electron thermal conductivity can be different. Here, since we focus on thermal conductivity, the average electron and phonon MFP is defined through the accumulation function (see Sec.S6) which describes the contributions of carriers with different MFPs to thermal conductivity. The final values of Λ are extracted at 50% of the accumulation function and the results are shown in Fig. 9. We also compare our calculated electron MFPs with the available reference data[29,66] as shown in the inset of Fig. 9. It should be noted that the MFP in the inset plot is calculated with the definition of Gall's[66] work to make a fair comparison. We can see that our calculated values agree well with the reference data. Importantly, it can be found that the phonon MFP is within 10 nm for all the 18 metals while the electron MFP ranges from 5 to 25 nm. Interestingly, we can see that the electron MFPs of nobles are larger than transition metals, which gives a general explanation of the reason why noble metals are good conductors. In addition, we can see that the MFPs of electron are in general larger than that of phonon, which indicates the electron thermal conductivity has a larger reduction and the phonon thermal conductivity is quite important in metal nanostructures. Furthermore, in order to understand the function of electron MFP in determining the electron transport properties, we plot the variation of electrical conductivity with electron MFPs and it is shown in Fig. 10. We can see that the materials with large electron MFPs generally have high electrical conductivity and the Pearson correlation between them is 0.90, which indicates that the transport properties of electrons are strongly correlated to the electron MFPs.

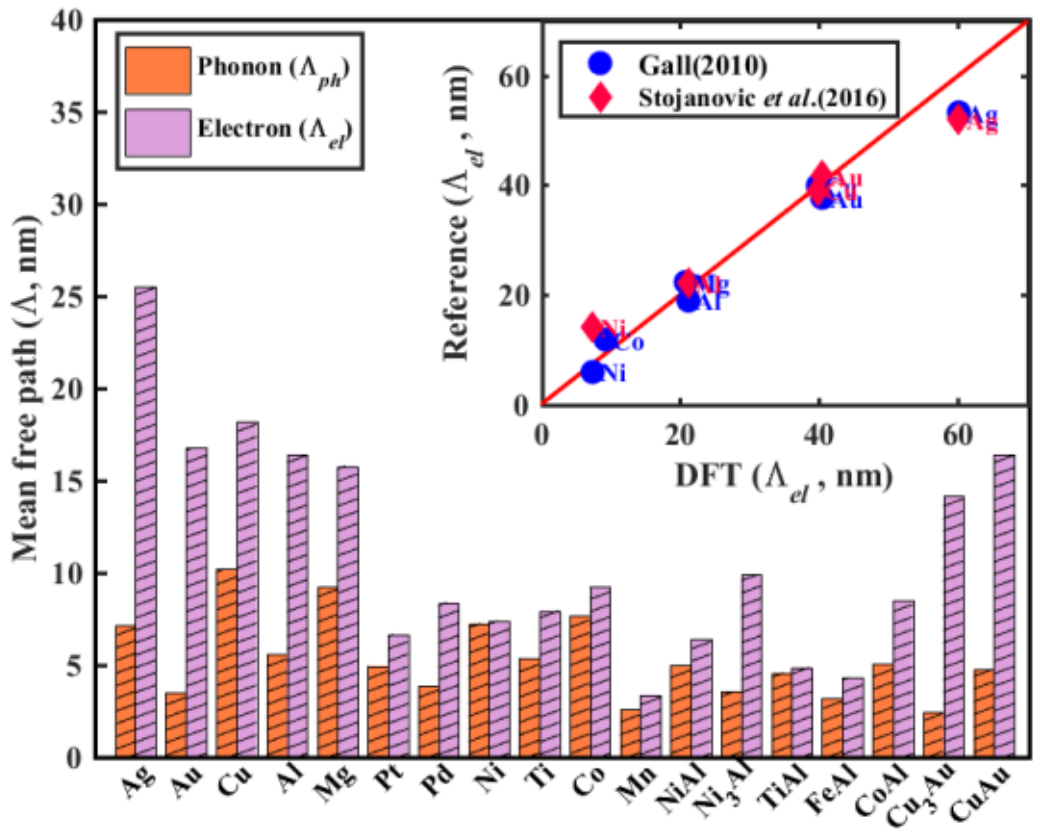

Fig. 9. Average phonon and electron mean free path of 18 metals at 300 K. The inset plot is the comparison of electron MFP between our calculated values and reference data[29,66].

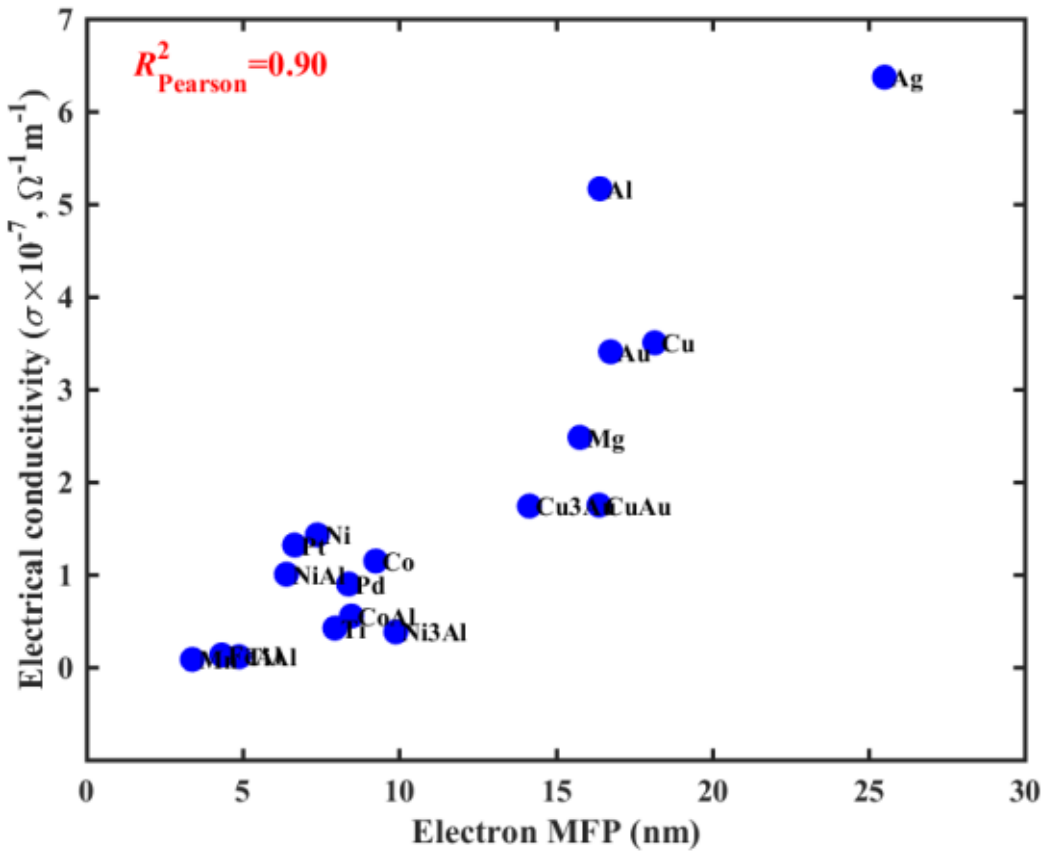

Fig. 10. Variation of electrical conductivity with electron MFPs for 18 metals at 300 K.

## 4. Summary

In summary, first-principles calculations are conducted to predict the mode-dependent thermal properties of 18 metals including noble metals, transition metals, alkali-earth metals, noble-intermetallic-compounds and transition-intermetallic-compounds at room-temperature. The first-principles predicted values of thermal conductivity and electrical conductivity agree well with experimental results. The first-principles data allows the quantification and the separation of the electron and phonon contributions to thermal conductivity. We find that phonon thermal conductivities which only consider phonon-phonon scattering are within a range of 2 - 30 W/mK, in which the phonon group velocity is the dominant factor of determining the phonon thermal conductivity. The phonon thermal conductivities will be reduced to 2 - 18 W/mK when the phonon-electron scattering is included, which finally results in the phonon thermal conductivity takes a proportion of 1% - 40% in total thermal conductivity. Moreover, we find that the electron-phonon coupling effect on phonon thermal conductivity in transition metals and intermetallic compounds is stronger than that of nobles, which is mainly due to the large electron-phonon coupling constant with a high electron density of states within Fermi window and high phonon frequency. In addition, it is found that noble metals hold very high electron thermal conductivity in range of 265 - 476 W/mK mainly due to weak electron-phonon coupling. Besides, the Lorenz ratios for all the 18 metals are calculated and it is found that there are larger deviations from the Sommerfeld value $L_0$=2.44×10$^{-8}$ WΩK$^{-2}$ in transition metals and TICs. Finally, it is shown that the MFPs for phonon (within 10 nm) are smaller than these of electron (5 - 25 nm). The long electron MFPs lead to large electrical and electron thermal conductivities in metals.

## Acknowledgement

This work was supported by the National Natural Science Foundation of China (Grant No. 51676121) and the Materials Genome Initiative Center of Shanghai Jiao Tong University. Simulations were performed at Center for High Performance Computing (π) of Shanghai Jiao Tong University. Zhen thanks for the financial support of Chinese Scholarship Council (CSC, 201806230169).